\documentclass[10pt]{iopart}
\usepackage{graphicx}
\usepackage{bm}
\usepackage{epstopdf}
\usepackage{subfigure}%
\usepackage{color}
\usepackage{lineno}
\usepackage[colorlinks,
linkcolor=blue,citecolor=blue,bookmarks,pdfstartview=FitH]{hyperref}
\usepackage{mathtools,amssymb,lipsum}

\usepackage{ulem}

\begin{document}
\title[EP transport in CoM space]{Energetic particles transport in constants of motion space due to collisions in tokamak plasmas} 

\author{Guo Meng$^{1}$, Philipp Lauber$^{1}$, Zhixin Lu$^{1}$, Andreas Bergmann$^{1}$, Mirelle Schneider$^{2}$
}
\address{ $1$ Max-Planck Institute for Plasma Physics, 85748 Garching, Germany \\
 $2$ ITER Organization, Route de Vinon-sur-Verdon, CS90046, 13067 St Paul-lez-Durance, France\\
}
\ead{guo.meng@ipp.mpg.de}
\vspace{10pt}
\begin{indented}
	\item[]\today
\end{indented}
\begin{abstract}

{The spatio-temporal evolution of the energetic particles in the transport time scale in tokamak plasmas is a key issue of the plasmas confinement, especially in burning plasmas. 
In order to include sources and sinks and collisional slowing down processes, a new solver, ATEP-3D was implemented to simulate the evolution of the EP distribution in the three-dimensional constants of motion (CoM) space. The Fokker-Planck collision operator represented in the CoM space is derived and numerically calculated. The collision coefficients are averaged over the unperturbed orbits to capture the fundamental properties of EPs. ATEP-3D is fully embedded in ITER IMAS framework and combined with the LIGKA/HAGIS codes. The finite volume method and the implicit Crank-Nicholson scheme are adopted due to their optimal numerical properties for transport time scale studies. ATEP-3D allows the analysis of the particle and power balance with the source and sink during the transport process to evaluate the EP confinement properties.}

\end{abstract}



\maketitle
\section{Introduction}

In tokamak plasmas, energetic particles (EPs) can be generated by fusion products, neutral beam injection, and other auxiliary heating with energy much greater than those of the background ions and electrons. 
The EPs can heat the plasma through Coulomb collisions with thermal particles, and improved confinements of EPs are needed for achieving high fusion gain. As the particle moves along its equilibrium trajectory in tokamak, its orbit can be described in the constants of motion (CoM) space \cite{zonca2015nonlinear}. The CoM generally used in gyrokinetic simulations of fusion plasma are particle energy $E$, toroidal canonical momentum $P_\zeta$, and magnetic moment $\mu$, which is determined by the magnetic field configuration. EPs' orbit can have a large radial excursion ($\sim 0.1 a$) and the finite orbit width effect is important to understand the EP transport. The CoM variables provide an appropriate way to describe the EP dynamics. Specifically, in full $f$ simulations the adoption of the CoM space is necessary to initialize a steady state distribution with the finite orbit width effects taken into account \cite{lu2022fullf}. A parametric equilibrium distribution function in the CoM space is applicable to gyrokinetic studies investigating the behavior of guiding centers with finite orbit width in axisymmetric tokamak plasmas, as discussed in \cite{di2012feq}.
The characteristic velocities of EPs are close to the phase velocities of Alfvén eigenmodes (AEs), which EPs can destabilize through resonant interactions\cite{todo2018RMP}. The AE induced EP transport is more efficient and faster than the collisional transport, as summarized previously \cite{heidbrink2020EPtransport}. 
However, collisions enhance the resonance width by moving particles in and out of the resonance islands  \cite{white2019collisionResBroad} and affect the mode evolution \cite{zhou2016collision}. On a longer timescale, the distribution function is shaped by sources, sinks, wave-induced transport, and collisions, which need to be considered in the EP transport model. 

The neoclassical transport and the bootstrap current induced by alpha particles and NBI were theoretically studied in Refs. \cite{hsu1992bootstrapAlpha,taguchi1996bootstrapNBI}, where was found that the bootstrap current induced by EP is about $10 \sim 20 \%$ of the background bootstrap current. The ion poloidal gyroradius $\rho_p$, which indicates the finite orbit width, can be comparable to or even larger than the characteristic length of the equilibrium $L_{eq}$, which consequently violates the assumption in the traditional neoclassical theory based on the asymptotic solution of the drift equation using $\rho_p$ \cite{hinton1976transport}. It has been studied that the small poloidal gyroradius assumption of standard neoclassical theory breaks down due to the wide orbit width analytically \cite{shaing1997ion} and numerically \cite{lin97PRL,bergmann2001Neo}. Indeed, the neoclassical transport in the large $\rho_p/L_{eq}$ regime has attracted attention due to its possible connection to the L-H transition \cite{helander1998bifurcated}.
In addition, the collision effect is important in the reduced transport models to study the long-timescale steady-state profiles. In SOLPS \cite{wiesen2015SOLPS-ITER} and ASTRA \cite{pereverzev2002astra}, the collision transport is simulated with coefficients predicted by neoclassical theory or matched to the experimental measurement.
The transport  of runaway electrons has been studied based on a fluid-kinetic code DREAM \cite{hoppe2021dream}.
For most codes using the Monte-Carlo collision operator, collisions represented in the velocity space $(v_\parallel,v_\perp)$  \cite{lin95pop} or the pitch angle scattering \cite{Boozer1981MC} are  adopted \cite{transp04,schneider2005SPOT}. Linearized collision operator in CoM coordinates $E,\mu$ is described and tested in \cite{xiong2008FPO}. Besides the collisional process, a general theoretical framework for the transport of Phase Space Zonal Structures (PSZS)  has been developed \cite{zonca2015nonlinear,falessi2019transport,Falessi2023IAEA,Zonca2023IAEA}. PSZS is the long-lived toroidal symmetric ($n=0$) structure that defines the nonlinear equilibrium in the presence of fluctuations such as Alfv\'enic instabilities. The PSZS theory is used to study the fluctuation-induced EP transport and in general includes collisional effects. Based on this framework, a reduced energetic particle (EP) transport code (named `ATEP') is developed. The primary objective of ATEP is to systematically encompass the intricate physics embedded in the overarching equations, as detailed in \cite{lauber2023IAEA,lauber2024NF}.  The general ATEP model addresses comprehensive physics, recovering critical gradient models\cite{waltz2014CGL}, the `kick' model\cite{podesta2014kick}, and quasi-linear resonance broadening models\cite{nikolaiRBQnf,meng2018NF} within appropriate limits. 

In this paper, we implement and investigate EP transport in the CoM space due to collisions. In order to include sources and sinks and collisional slowing down processes, a new solver, ATEP-3D is implemented for the energetic particle transport modeling on the transport time scale. The collision operator is represented in the three-dimensional (3D) CoM space and averaged over the unperturbed orbits. The collision frequencies are given by the sum of the contributions from different sources. The finite volume method (FVM) and the implicit scheme are adopted for optimized numerical properties.
ATEP-3D allows the analysis of the particle and power balance with the source and sink during the transport process to evaluate the EP confinement properties.
The following part of this article is organized as follows. The physics model and the numerical schemes are givens in Sections \ref{sec:physics} and \ref{sec:numerics} respectively. The description of the experimental case and the simulations results are represented in Section \ref{sec:results}, followed by  Section \ref{sec:conclusions} as the conclusion part. 

\section{Physics models}
\label{sec:physics}
The evolution equations for PSZS are derived in a recent work, which describe the slow evolution of macroscopic plasma profiles and the more rapid phase space corrugations on meso-scales due to fluctuations in toroidal geometry \cite{falessi2019transport}. The nonlinear gyrokinetic theory is utilized to define a neighboring nonlinear equilibrium consistent with toroidal symmetry-breaking perturbations. It introduces PSZS as a means to capture the fluctuation-induced corrugations of the reference state \cite{falessi2019transport}. The collisional transport and its synergistic interplay with fluctuation-induced transport are important to EP transport studies. Consequently, the focus of this work is to derive the collisional transport in the CoM space. By exploring this aspect, we aim to gain a better understanding of how collisions influence the transport of EPs and how they interact with the transport induced by fluctuations. This analysis will contribute to a comprehensive understanding of the complex mechanisms governing EP transport in fusion plasmas. Previously, in the CGM (critical
gradient model)\cite{waltz2014CGL} a 1D radial diffusion coefficient was used with a recipe for the onset of transport (the growth rate of the toroidal Alfv\'en eigenmode $\gamma_{\rm TAE}> E\times B$  shearing rate) and an ad  hoc saturation rule assuming a diffusion coefficient $D=1\sim10 \rm{ m}^2/s$ \cite{zou2022CGM}. No phase space dependence is considered, with the assumption of diffusive transport, despite sufficient evidence of non-diffusive EP transport in theory and experiment.
The kick model \cite{podesta2014kick} resolves the phase space dynamics, but the self-consistent loop between the mode amplitude, the evolution of EP distribution, and background profiles is not taken into account. More fundamentally, the kick model is a probabilistic approach, while the PSZS theory solves transport equations in phase space. However, the comparison is an important step for model validation in the future \cite{lauber2024NF}.
The RBQ (Resonance Broadened Quasilinear) model is built upon the foundation of the quasilinear and diffusive model, expanding to include the broadened resonance inherent in wave-particle interactions. The RBQ code is reduced into a 1D $P_\zeta$ \cite{nikolaiRBQnf} and a 2D $(E, P_\zeta)$ \cite{gorelenkov2019RBQ2D} CoM space. Moreover, collisions are simplified into a diffusive collision operator.
Our model aims to capture the 3D phase structure of EPs and the nonlinear interactions between EPs, the background, and the mode evolution.
The distribution function of EPs in the CoM space is modeled in the form
\begin{equation} \label{eq:DF}
    \frac{\partial f}{\partial t}= C(f)+ \gamma f+S,
\end{equation}
where $f({\bm X},t)$ is the EP distribution, ${\bm X}$ denotes the CoM phase space coordinates, the $C(f)$ is the collision operator, and $\gamma(\bm{X},t)$ and $S({\bm X},t)$ are the sink/source terms. In tokamaks, an unperturbed particle trajectory is periodic in three canonical angles and can be described by the three CoM: particle energy $E$, toroidal canonical momentum $P_\zeta$ and magnetic moment $\mu$. We choose ${\bm X}=(P_\zeta,E,\Lambda)$, where $\Lambda=\mu B_0/E$ is also a constant of motion.  

\subsection{Collision operator in $(P_\zeta,E,\Lambda)$}
 The linearized collision operator appeared in early literature as the approximate Fokker-Planck collision operator for the transport theory applications \cite{hirshman1976}.
 The CoM orbit representation is especially useful for treating EPs that have large orbits \cite{rome1979topology}. 
 The neoclassical transport studies considering large orbits width is derived \cite{shaing1997ion}. 
 Transport fluxes formulated using CoM first appeared in \cite{zaitsev1993three} and later in \cite{hirvijoki2013monte}. 
In this work, we derive the linearized collision operator in the $(P_\zeta,E,\Lambda)$ CoM coordinates. 
We assume that the energetic particles generated by NBI or fusion processes are immediately ionized, and we take into account the Coulomb collisions between charged particles. These EPs exhibit azimuthal symmetry with respect to the magnetic field. The collision can be derived from the Rosenbluth expression\cite{rosenbluth1957fokker}. Inter-species EP collisions are neglected due to their high energy and low density. The collisions experienced by EPs primarily involve interactions with the background thermal ions and electrons, which can be described by a Maxwellian distribution. In our work, a linear collision operator in the CoM space is adopted. The collision operator of the guiding center distribution function in the velocity 
space coordinate $(v_\parallel,v_\perp^2)$ is formulated by the Eq. (15) of \cite{lin95pop}, 
\begin{equation} \label{eq:2}
       \begin{aligned}
        C(f)=&  \frac{\partial}{\partial v_\parallel}(\nu_{s\parallel}f) + \frac{\partial}{\partial v_\perp^2}(\nu_{s\perp}f) \\
                & + \frac{\partial^2}{\partial v_\parallel \partial v_\perp^2}(\nu_{\parallel\perp}f) +\frac{1}{2} \frac{ \partial ^2 } {\partial v_\parallel ^2} (\nu_\parallel f) + \frac{1}{2}\frac{\partial^2}{(\partial v_\perp^2)^2} (\nu_{\perp}f),
        \end{aligned}   
\end{equation}
where the first two terms of the Right Hand Side (RHS) are the slowing down terms in $(v_\parallel,v_\perp^2)$ directions respectively, and the last three terms are the diffusion. The collision coefficients are
\begin{equation}
 \begin{aligned}
        &\nu_{s\parallel}=v_\parallel F, \;\;
        \nu_{s\perp}=2v_\perp^2 F-v_\perp^2 H -(2v_\parallel^2+v_\perp^2)G,\;\;\\
        \nu_\parallel=v_\parallel ^2 H+v_\perp^2 G,\;\;
        &\nu_\perp=4v_\perp^2(v_\perp^2 H+v_\parallel^2 G),\;\;
        \nu_{\parallel \perp}=2v_\perp^2 v_\parallel(H-G).
\end{aligned}   
\end{equation}
Functions $F$, $G$ and $H$ are defined by 
\begin{equation}
 \begin{aligned}
        &F=\left( 1+\frac{m_{EP}}{m_\beta} \right)\phi(x)\nu_0, \;\;
        G=\left[ \left( 1-\frac{1}{2x}\right)\phi(x)+\frac{d\phi(x)}{dx} \right]\nu_0,\;\;
        H=\frac{1}{x}\phi(x)\nu_0.
\end{aligned}   
\end{equation}
respectively. Here $x=v^2/v^2_{th \beta}$ and $\phi(x)$ is the Maxwellian integral defined by
$
    \phi(x)=\frac{2}{\sqrt{\pi}}\int_0^x e^{-t} \sqrt{t} \; dt.
$
Here, $m_\beta$ and $v_{th\beta}=\sqrt{2k_BT_\beta/m_\beta}$ represent the mass and thermal velocity of the background Maxwellian-distributed particles.
The basic collision frequency of EP colliding with background $\beta$ particle here is defined by
\begin{equation}
    \nu_0=\frac{ n_\beta q^2_{EP}q_\beta ^2 \ln \Lambda_{EP,\beta}}{4\pi \epsilon_0^2 m_{EP}^2 v^3},
\end{equation}
where $\ln \Lambda_{EP,\beta}$ is the Coulomb logarithm, $\epsilon_0$ is the vacuum permittivity.
Due to the constraint that the particle's perpendicular energy must always be smaller than the total energy if utilizing coordinates such as $E$ and $\mu$, the CoM space tends to have a significant portion where $f$ vanishes, namely $f(\mu>E/B)=0$. This issue necessitates specific treatment, such as employing triangular meshes, to appropriately handle the loss cone boundaries \cite{xiong2008FPO}. In order to simplify the numerical simulation process, an alternative coordinate, $\Lambda$, is chosen instead of $\mu$. Here, $\Lambda$ is defined as $\Lambda=\mu B_0/E$, which is a dimensionless quantity that provides a more convenient way to describe the equations. The three CoM are \cite{pinches1998hagis,white2013TCP}
\begin{equation}
 \begin{aligned}
        &P_\zeta=mgv_\parallel/B+q\psi,\\
        &E=m(v_\parallel^2+v_\perp^2)/2,\\
        &\Lambda=\frac{B_0}{B}\frac{v_\perp^2}{v_\parallel^2+v_\perp^2}.\\
\end{aligned}   
\end{equation}
with $\psi$ is the poloidal flux coordinate, $g$ is an equilibrium function. In HAGIS code, Boozer coordinates are used. The covariant form of the magnetic field is $B=g(\psi)\nabla \zeta+ I(\psi) \nabla \theta+ \delta(\psi,\theta)\nabla\psi$, where $g$ is the covariant field component in $\zeta$ direction and $g$ is a function of $\psi$ \cite{pinches1998hagis,white2013TCP}. Then we can derive
\begin{equation} \label{eq10}
 \begin{aligned}
       &\frac{\partial}{\partial v_\parallel}=m\sigma\sqrt{\frac{2E}{m}\left( 1-\Lambda \frac{B}{B_0}\right)} \frac{\partial}{\partial E}
        -\sigma \Lambda \sqrt{\frac{2m}{E}\left( 1-\Lambda \frac{B}{B_0}\right)}  \frac{\partial}{\partial \Lambda} + \frac{mg}{B}\frac{\partial}{\partial P_\zeta},\\
        &\frac{\partial}{\partial v_\perp^2}=\frac{m}{2} \frac{\partial}{\partial E}+\frac{m}{2E}\left( \frac{B_0}{B}-\Lambda \right)\frac{\partial}{\partial \Lambda},\\ 
\end{aligned}  
\end{equation}
where $\sigma=sign(v_\parallel)$.
Substitute Eq. \ref{eq10} to Eq. \ref{eq:2} we can derive the collision operator in CoM space. The collision operator is formulated 
\begin{equation}
\label{eq:coll}
    C(f)=\frac{\partial}{\partial P_\zeta}  (D_{P} f)+ \frac{\partial }{\partial E} (D_E f) +  \frac{\partial }{\partial \Lambda} (D_\Lambda f) + \frac{\partial^2 }{\partial P_\zeta^2} (D_{PP}  f) + \frac{\partial^2 }{\partial E^2}(D_{EE} f) + \frac{ \partial ^2 } {\partial \Lambda ^2}(D_{\Lambda\Lambda} f).
\end{equation}
\begin{equation}
    \begin{aligned}
        &D_{P}=\frac{mg}{B}\nu_{s\parallel}, \;\;
        D_E=mv_\parallel\nu_{s\parallel}+\frac{m}{2}\nu_{s\perp},\;\;
        D_\Lambda=-2\Lambda \frac{v_\parallel}{v^2}\nu_{s\parallel}+\frac{1}{v^2}\left(\frac{B_0}{B}-\Lambda\right)\nu_{s\perp},\\
        &D_{PP}= \frac{m^2g^2}{2B^2} \nu_\parallel,\;\;
        D_{EE}=\frac{m^2v_\parallel }{2}\nu_{\parallel \perp}+\frac{m^2v_\parallel^2}{2} \nu_\parallel+\frac{m^2}{8}\nu_{\perp},\\
        &D_{\Lambda\Lambda}=-2\Lambda\frac{v_\parallel}{v^4}\left(\frac{B_0}{B}-\Lambda\right)\nu_{\parallel\perp}+2\Lambda^2\frac{v_\parallel^2}{v^4}\nu_\parallel+\frac{1}{2v^4}\left(\frac{B_0}{B}-\Lambda\right)^2\nu_\perp.\\
\end{aligned}   
\end{equation}
The collision coefficients used in Eq. \ref{eq:coll} are averaged over unperturbed orbits. The leading terms are kept in the equations and the off-diagonal diffusion terms are omitted. Additionally, the derivation in the Appendix substantiates the absence of off-diagonal diffusion terms. The diffusion coefficient in $\Lambda$ direction $D_{\Lambda\Lambda}$ is zero when $\Lambda=0$. To include the finite orbit width effect, an average over the unperturbed orbit is adopted as follows,
\begin{equation}
  \left<D\right>=\int D \; dt /\tau_b
\end{equation}
where $\tau_b$ is the bounce time, during which particles complete their poloidal orbits once, i.e. $\tau_b=\oint\frac{1}{\dot\theta} d\theta$ with the range of integration for a trapped particle being $(-\theta_b,\theta_b)$ and for a passing particle being $(0,2\pi)$, $\theta_b$ is the bounce angle. For simplicity, the average $\left<\right>$ is omitted. The numerical calculation of collision coefficients in the CoM space is detailed in Section \ref{sec:results}.

\section{Numerical implementation}
\label{sec:numerics}
\subsection{Code structures}
The transport equation Eq. \ref{eq:DF} is solved in the ATEP-3D. The spatial differential is discretized using the finite volume method (FVM), a widely used technique in fluid simulation due to its favorable conservation properties. Specifically, a fully implicit Crank-Nicholson scheme is employed in order to ensure the accurate representation of the long-term behavior of EP transport. In the boundary, the finite difference boundary condition is used with different choices. The ATEP-3D numerical solver has been implemented using MATLAB, and a Fortran version is currently under development. To facilitate efficient testing of conservation properties and steady state studies, 1D and 2D solvers are utilized. The ATEP-3D code follows an object-oriented programming paradigm, consisting of two classes: {\it {atep\_3d}} which serves as the core solver, and {\it {dist\_exp\_3d\_cls}} an Experimental Data Interface.

\subsection{Finite volume method for spatial discretization}
The Finite Volume Method operates by subdividing the simulation domain into discrete cells and solving the equation at the centroid of each cell. The simulation domain is divided into $N_{P}\times N_E\times N_\Lambda$ cells with center values denoted using integer indices $(i,j,k)$. The center values within the cells are denoted using integer indices $(i,j,k)$, while the points located on the cell faces are indicated by indices with half integers. Since the advection terms in the equations can be negative and positive, the centered discretization scheme is used. The phase-space derivatives of the partial differential equation is evaluated at the cell face,
$
    \frac{\partial f}{\partial x^m}=\frac{f_{i^m+1/2}-f_{i^m-1/2}}{\Delta x_i^m},
$
where $x^m$ is the $m$th coordinate and $i^m$ is the index in $x^m$ direction of this cell.
Since the advection terms in the equations can be negative and positive, the centered discretization scheme is used. The phase-space derivatives are discretized using a linear centered scheme, i.e., the value of a quantity $X$ at the cell face is taken as the average of the adjacent cells, 
$
    X_{i\pm\frac{1}{2}}=\frac{1}{2}\left(X_i+X_{i\pm 1}\right).
$
The transport coefficients are dependent on the phase-space coordinates but remain constant over time.

\subsection{Implicit scheme for time evolution}
The time derivatives are discretized using the Crank-Nicholson method,
\begin{equation}
    \frac{f^{l+1}-f^l}{\Delta t}=\frac{1}{2}\left({\text {RHS}} ^{l+1}+{\text {RHS}}^{l}\right),
\end{equation}
where $l$ represents the time step index. Currently, a uniform grid and a fixed time step size are employed. 

For simplicity, the source is assumed $S^l\approx(S^l+S^{l+1})/2$. An explicit source term is utilized. The transport coefficients $D_x,\; D_{xx}$ are time-invariant.
The derivatives appearing in the diffusion terms are also conveniently discretized using a central difference approximation.
Move all terms at time step $l+1$ in to the left hand side and those at step $l$ to the right hand side, we obtain a system of linear equations.
\begin{equation}
    \Bar{\Bar{M}}^*_{LHS}\cdot \Bar{F}^{l+1}=\Bar{\Bar{M}}^*_{RHS}\cdot \Bar{F}^{l}+\Bar{S}^l
\end{equation}
where $\Bar{F}$ and $\Bar{S}$ are arrays and $\Bar{F}_{i+(j-1)N_P+(k-1)N_PN_E}=f_{ijk}$. The degree of freedom of this system is $N_{dof}=N_{P}\times N_E\times N_\Lambda$. The matrix representation of a single time step in the process involves the differential operator matrix $\Bar{\Bar{M}}^*$, which is a square matrix with size $N_{dof}\times N_{dof}$. It should be noted that the matrices $\Bar{\Bar{M}}^*$ are constructed based on selected boundary conditions and are independent of time.
\subsection{Numerical Verification with Varied Boundary Conditions}
Manufactured solutions are employed to assess the solver's performance, such as the pure advection and diffusion equations. Throughout our investigation, we investigate the mass conservation and identify the crucial role of boundary conditions in ensuring conservation.
\begin{figure}[htbp]\centering
    \centering
    \includegraphics[width=0.8\textwidth]{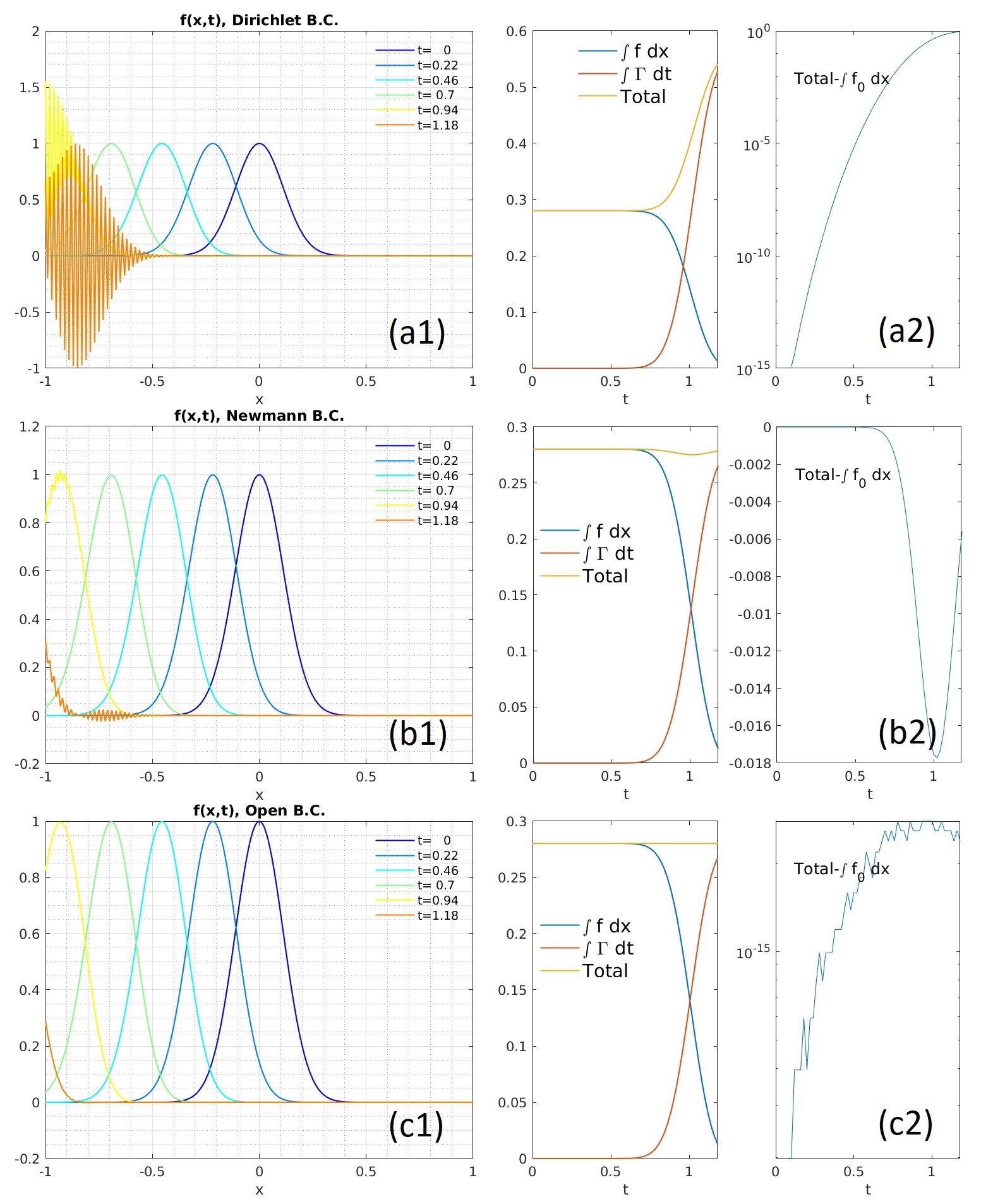}
    \caption{Comparison of different boundary conditions. (a) Dirichlet (b) Neumann (c) Open}
    \label{fig:1D}
\end{figure}
This section presents the results and discussions regarding the advection in the transport equation. The advection terms, such as the slowing down of energy due to collisions, induce a shift in the distribution function $f$ while preserving its shape. A positive advective coefficient results in a negative flow. The EPs experience deceleration due to interaction with the background. In the transport model, EPs at the low energy tail is considered to merge to background ions, leading to  EP ``losses" at the boundaries. To address this, an ``open" boundary condition is implemented, allowing the flux to pass through the boundary without distorting the distribution function or affecting the interior solution. The open boundary condition is represented by the equation: $\partial^2 f/\partial x^2=0$. In this section, a 1D test case is utilized to elucidate the numerical calculation of advection and the significance of open boundary conditions,
\begin{equation}
\label{eq:1d}
    \frac{\partial f}{\partial t}= \frac{\partial}{\partial x} (D_x f) +\frac{\partial^2}{\partial x^2} (D_{xx} f).
\end{equation}
The advective coefficient is assumed to be constant, and the initial distribution follows a Gaussian function. When $D_x=1, D_{xx}=0$ and the initial distribution $f_0=\exp({-x^2/0.025})$, the analytical solution of Eq. \ref{eq:1d} is $f=\exp[-(x+D_x t)^2/0.025]$ with free boundaries. Different boundary conditions are implemented and tested. Since the simulation domain is not infinite, when the distribution function passes through the boundary the numerical error will arise rapidly due to the Dirichlet and Neumann boundary conditions. With using the open boundary condition, the conservation is satisfactory and the flux out of the system is properly calculated. As shown in Fig. \ref{fig:1D}(a), the left figure shows the result using zero boundary condition. When the time is longer than slowing down time, the numerical error increases at the lower boundary.
The Neumann boundary condition is better but the conservation of total particle number is not ensured as shown in Fig. \ref{fig:1D}(b).
With the open boundary condition, the numerical error is negligible and the particle number is conserved to machine precision as shown in Fig. \ref{fig:1D}(c2). The Courant number $C=\frac{D_x\Delta t}{\Delta x}=\frac{1\times0.02}{0.01}=2$ in these cases, and the numerical results have converged.

\section{Simulations results}
\label{sec:results}
\subsection{Orbit averaged collision in the CoM space}

The collision coefficients are numerically calculated by the HAGIS code. The neoclassical transport of particles with wide orbits has been studied and validated \cite{bergmann2001Neo}. To obtain collision coefficients in the CoM space, the implementation involves following test particles in the equilibrium and then averaging the coefficients over one poloidal orbit.  Calculation of collision coefficient is added in the IMAS \cite{imbeaux2015IMAS} versions of HAGIS and related wrapper module (FINDER). And the averaged diffusion coefficients are stored in distribution IDS. 
\subsection{ITER pre-fusion discharge}
\label{sec:setup}
\begin{figure}[htbp]\centering
    \centering
    \includegraphics[width=0.48\textwidth]{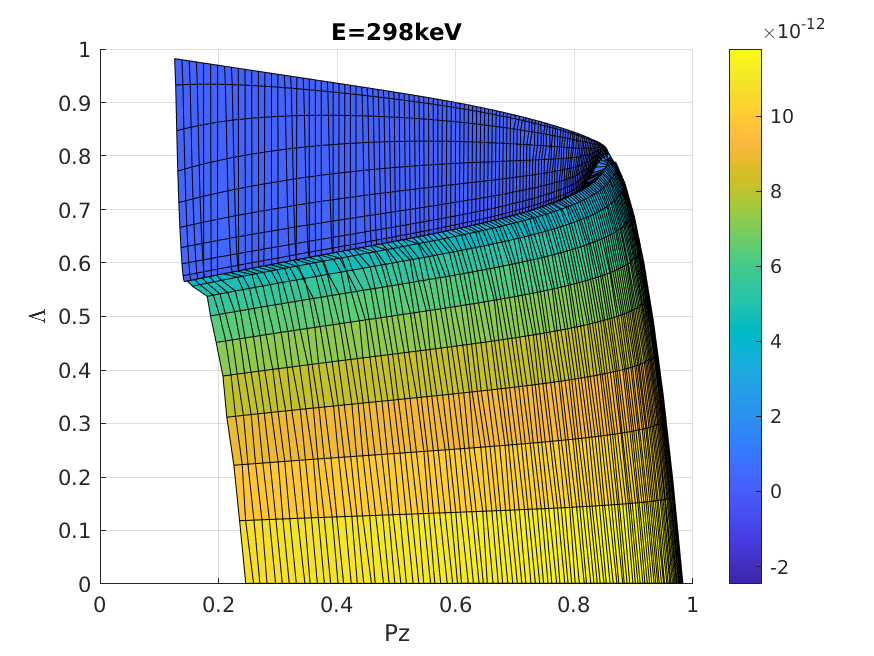}
    \caption{Collisional transport coefficient $D_{P}$ of $E=298 \;KeV$ particles in the constants of motion space.}
    \label{fig:nus1}
\end{figure}

In this work, we used the ITER pre-fusion power operation (PFPO) discharge (shot number 100015, run number 1) \cite{schneider2019}. The PFPO phases are featured by low magnetic field, plasma current and density (1.8 T and 5 MA for a density between $40\%$ and $50\%$ of the Greenwald density).
In the following, we adopt the equilibrium and other variables from the ITER PFPO discharge for the calculation of the EP collisions and the transport. 
The collision coefficients vary along particle orbits and the effects of the finite obit width and variation of the background profiles are important. In this work, uniform background density and temperature profiles of thermal ion and electron are assumed. The averaged collision coefficients in the CoM space for this case are calculated. In Fig. \ref{fig:nus1}, the collision coefficient $D_{P}$ of co-going particles with $E=298 \;keV$ is shown for example. For the co-passing particles, the $D_{P}$ is positive, it is due to the velocity drag. The $D_{P}$ of passing particle is larger since it is related to the parallel velocity. For trapped particles, $D_{P}$ is small. This illustrates that the collision coefficients depend on the particles orbit types, which can be described in the CoM space. All collision coefficients in the CoM space are constructed in this way. 

\subsection{Collisional EP transport}
\begin{figure}[htbp]\centering
    \centering
    \includegraphics[width=0.8\textwidth]{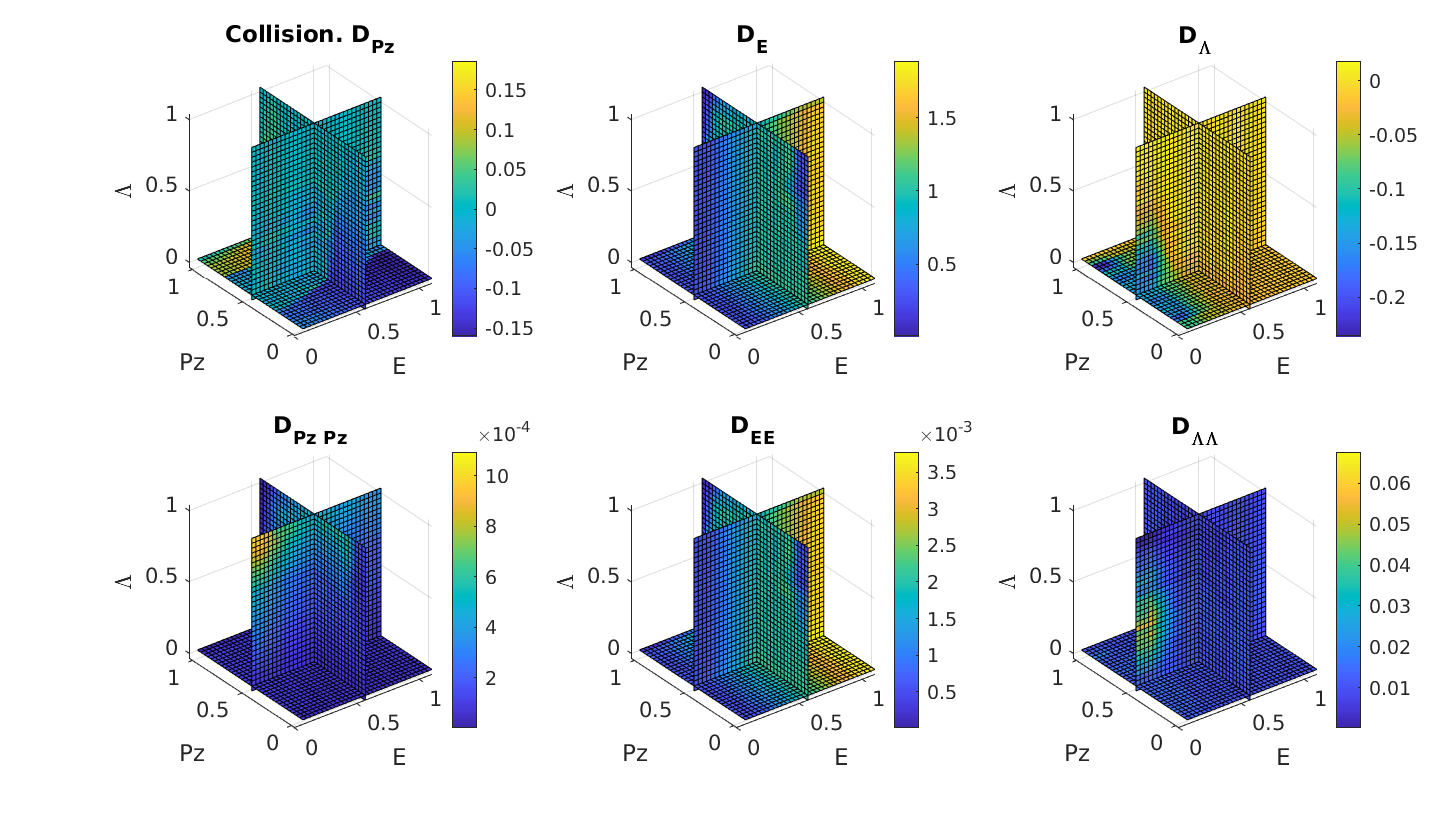}
    \caption{Collision coefficients in the three dimensional constants of motion space. The collision coefficients are normalized according to the  normalization of energy, $P_\zeta$ and time. }
    \label{fig:collCOM3D}
\end{figure}
\begin{figure}[htbp]\centering
    \centering
    \includegraphics[width=0.8\textwidth]{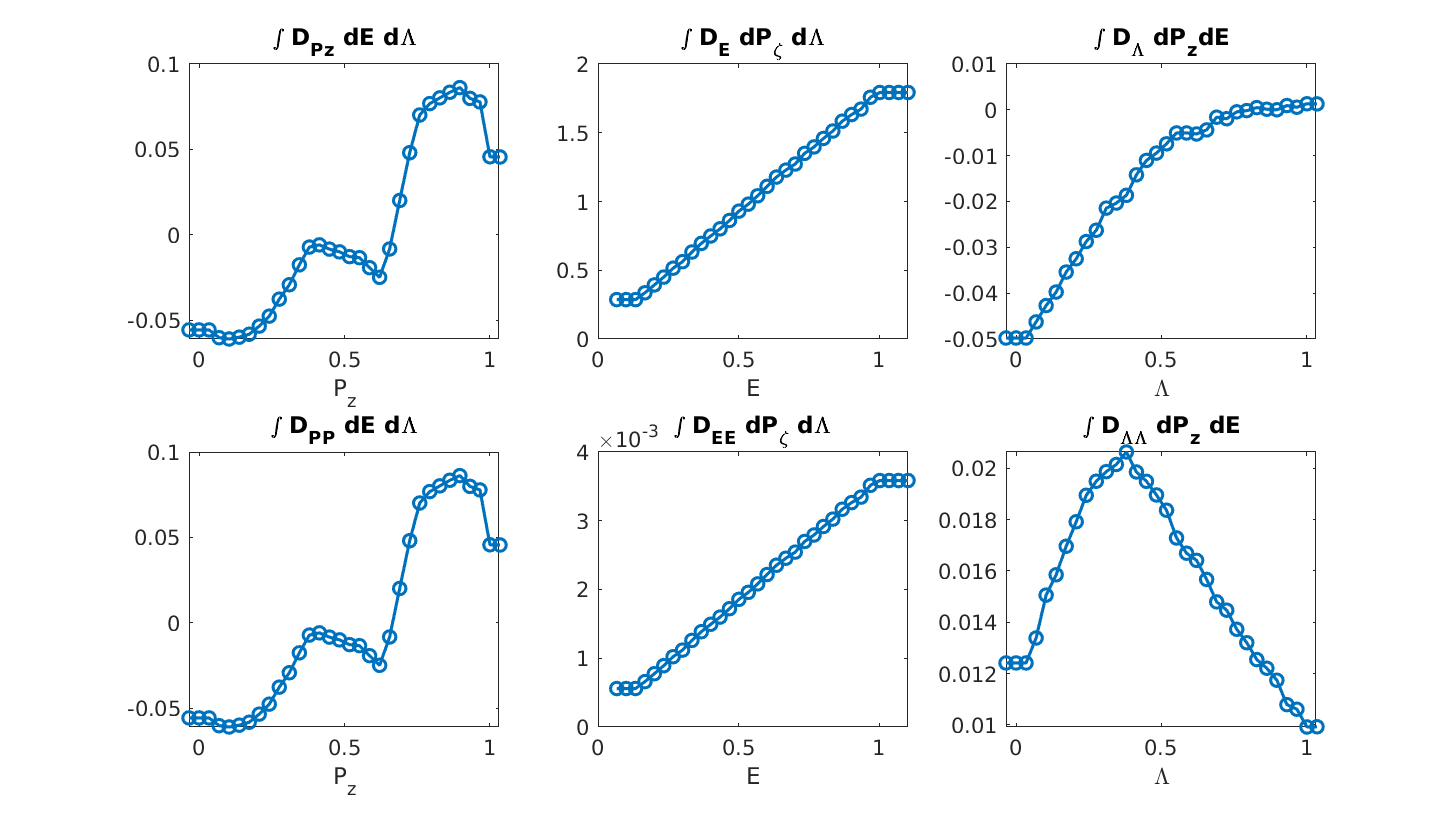}
    \caption{Integrated collision coefficients.}
    \label{fig:collision1d}
\end{figure}
\begin{figure}[htbp]\centering
    \centering
    \includegraphics[width=0.85\textwidth]{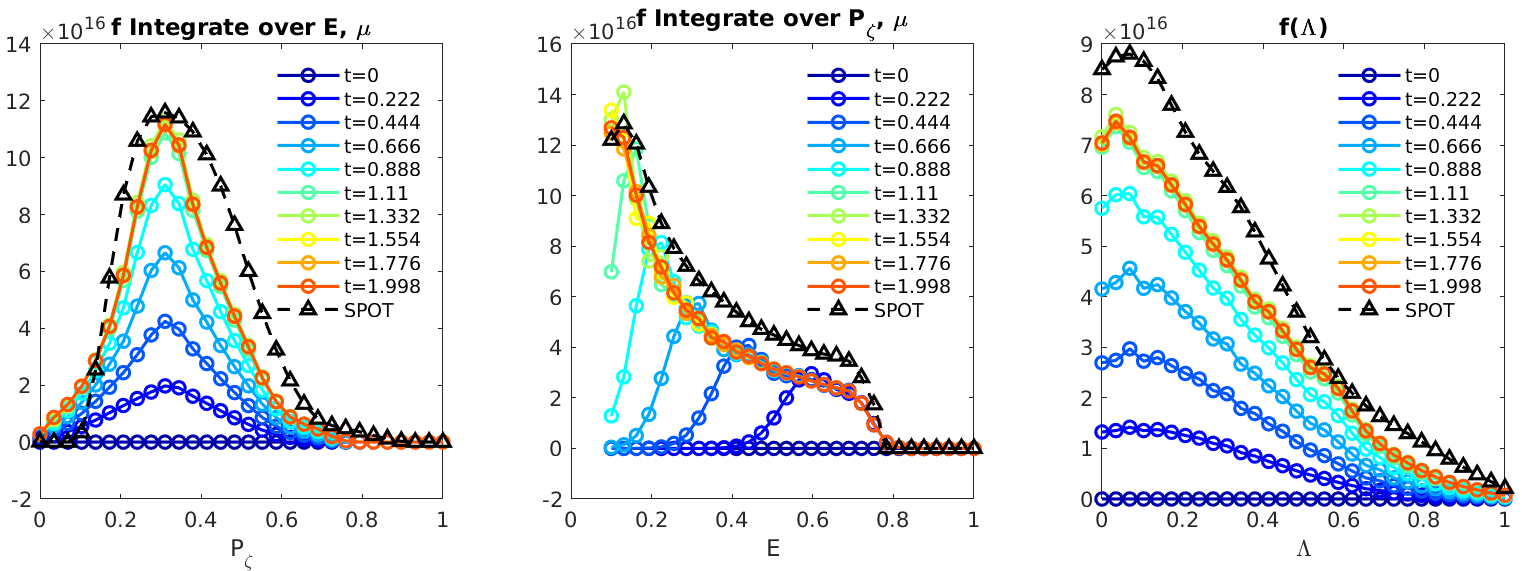}
    \caption{The figure shows the time evolution of the EP distribution simulated by ATEP-3D, represented by the color lines with circle markers. The black dashed line with triangle markers represents the SPOT result.}
    \label{fig:steadystate3d}
\end{figure}

In the following, we show the results of the NBI slowing down process simulated by ATEP-3D. The simulation regime of energy is $0.1\; MeV$ to $1\; MeV$. In this work, we are more concerned with the slowing down process of EPs and thus we use $0.1\;MeV$ as the lower boundary along $E$. In the future we will use a similar rule as TRANSP \cite{ TRANSP}, that treat all particles below  $1.5v_{ \rm{thermal}}$ as the main thermal species. In addition, at the even lower energy range, as the EPs merge to the background ions, the evolution of the background ion also needs to be modeled, which merits more effort in our future work.
In simulations, the energy is normalized to $1\;MeV$. 
The normalization of $P_\zeta$ and $\Lambda$ is chosen so that the normalized values are from 0 to 1. The time unit is second. The collision coefficients are normalized accordingly. 
The three-dimensional structure of the normalized collision coefficients are shown in Fig. \ref{fig:collCOM3D}. The advection along energy direction ($D_E$) is the dominant one indicated by its magnitude. The advection along $P_\zeta$ is contributed by both co-passing and counter-passing particles, corresponding different signs in $D_{P}$. The diffusion terms are dominated by $D_{\Lambda\Lambda}$.
The 1D integrated normalized collision coefficients are shown in Fig. \ref{fig:collision1d}, giving the structure and magnitude of the advection and diffusion coefficients in all directions.
In order to model the process of the EP evolution with source, the NBI source is modeled by a Gaussian function, $$S=A_s\exp\left\{-\left[(P_\zeta-0.31)^2/\delta P^2_\zeta+(E-0.745)^2/\delta E^2+(\Lambda-0.1)^2/\delta \Lambda^2 \right]\right\},$$ where $A_s=2\times10^{18},\delta P_\zeta=0.2,\delta E=0.05,\delta \Lambda=0.4$.  The initial distribution is $f_0=0$. 
Steady-state of EP distribution due to collisions and source is reached at about $1 s$. Clearly, the main processes are energy slowing down and pitch angle scattering.
We compare the NBI case with NEMO/SPOT. NEMO is to simulate the NBI deposition. SPOT (simulation of particle orbits in a tokamak) is an orbit following Monte Carlo code with Fokker-Planck collision operator \cite{schneider2005SPOT}. The birth energy is $745\;KeV$ for both ATEP-3D and NEMO/SPOT, without the consideration of the half or third energy of EPs since ITER will employ a neutral beam injector with negative ion source. The SPOT result is shown by the black line in Fig. \ref{fig:steadystate3d}. The steady-state solution of ATEP-3D agrees reasonably well with SPOT result. The discrepancies may come from that SPOT using the realistic density and temperature profiles. Additionally, the physics model and simplifications in NEMO/SPOT are not identical to those used in ATEP-3D, such as the parametrization of source and sink terms. Further detailed comparisons with other transport codes will be carried out in the future.
The back-mapping of EP distribution in CoM space to the real space will be performed as described in \cite{lauber2023IAEA}.


\section{Conclusions}
\label{sec:conclusions}
This study focuses on the transport of energetic particles in tokamak plasmas. Due to the relatively large orbit width resulting from the high energy of the energetic particles, the collision operator is expressed in constants of motion space and averaged along the unperturbed particle trajectory, which yields the diffusion and drag coefficients in constants of motion space. 
The numerical tool ATEP-3D has been developed, using the finite volume method and implicit Crank-Nicolson scheme. Favorable numerical behaviors are demonstrated in terms of conservation properties, large time step size and long time scale numerical behavior. The open boundary condition is shown to be a key ingredient for the modelling of the transport problem in the presence of source and sink as well as the collision terms.
As the numerical verification, the numerical results in one dimensional case show good agreement with the analytical results.
The ITER prefusion discharge serves as a test case, demonstrating the ATEP-3D code's capability in three-dimensional scenarios. The collisional coefficients in the constants of motion space are calculated and the slowing-down process of the EP transport is simulated. Adding the transport process due to Alfv\'enic perturbations \cite{lauber2023IAEA}, the ATEP-3D code will be applied to the studies of the evolution of the EP distributions with the  phase space zonal structure taken into account in a consistent way.

\section*{Appendix A. Collision operator in $(E,\Lambda)$ space}

Start from page 4 of Rosenbluth's paper \cite{rosenbluth1957fokker}, and substitute the coordinates $E,\Lambda,\phi$, where $E=v^2/2,\Lambda=\mu/E,\mu=v^2_\perp/2B$, $\phi$ is the gyro-angle. The mass unit is 1 and the unit of $B$ is $B_0$. Then we have,
\begin{equation}
\begin{aligned}
    & q^1=E, \quad    q^2=\Lambda, \quad   q^3=\phi,\\
    & ds^2=\frac{1}{2E}dE^2+\left[ \frac{BE}{2\Lambda}+\frac{BE}{2(1/B-\Lambda)}\right] d\Lambda^2+2EB\Lambda d\phi^2 ,\\
    &a_{11}=\frac{1}{2E},\; a_{22}=\frac{BE}{2}\frac{1}{\Lambda(1-B\Lambda)}, \; a_{33}=2EB\Lambda,\\
    &\qquad\qquad \qquad \qquad\qquad \qquad  a_{ij}=0 \quad \text{if }  i \neq j,  \\
    &a^{11}=2E, \; a^{22}=\frac{2}{BE}\Lambda(1-B\Lambda), \; a^{33}=\frac{1}{2EB\Lambda}\\
    &\qquad\qquad \qquad \qquad \qquad \qquad  a^{ij}=0 \quad \text{if }  i \neq j,  \\
    & a=\det(a_{\mu\nu})=\frac{B^2 E}{2(1-B\Lambda)}, \quad a^{\frac{1}{2}}=\frac{BE}{v_\parallel}, \\ 
\end{aligned}
\end{equation}
where $v_\parallel=\sigma\sqrt{2E\left( 1-\Lambda B\right)}$, $\sigma=sign(v_\parallel)$.  We can derive 
\begin{equation}
\begin{aligned}
    &T_a{}^1=2E\frac{\partial h}{\partial E}, \quad  T_a{}^2=\frac{2\Lambda(1-B\Lambda)}{BE} \frac{\partial h}{\partial \Lambda},\quad T_a{}^3=0,\\
    &(f T_a{}^\mu)_{,\mu}=\frac{v_\parallel}{BE}\frac{\partial}{\partial E}\left(\frac{2BE^2}{v_\parallel} f\frac{\partial h}{\partial E} \right)+\frac{v_\parallel}{BE}\frac{\partial}{\partial \Lambda} \left[ \frac{2\Lambda(1-B\Lambda)}{v_\parallel} f \frac{\partial h}{\partial \Lambda}\right]
\end{aligned}
\end{equation}

The second-rank tensor $S^{\mu \nu}$ in the $E,\Lambda,\phi$ coordinates follows 
\begin{equation}
   \begin{aligned}
    & S^{11}=4E^2 \frac{\partial^2 g}{\partial E^2}+2E\frac{\partial g}{\partial E}\\
    & S^{12}=4\Lambda \left( \frac{1-B\Lambda}{B} \right) \left( \frac{\partial^2 g}{\partial E\partial \Lambda }-\frac{1}{2E}\frac{\partial g}{\partial \Lambda} \right) \\
    & S^{13}=S^{23}=0\\
    &S^{22}=\left[ \frac{2\Lambda(1-B \Lambda )}{BE}\right]^2 \left[\frac{\partial^2 g}{\partial \Lambda^2 }+\frac{BE}{2\Lambda(1-B \Lambda )}\frac{\partial g}{\partial E} - \frac{2\Lambda B -1}{2\Lambda(1-B \Lambda )}\frac{\partial g}{\partial \Lambda} \right]\\
    &S^{33}=\frac{1}{2EB\Lambda}\left( \frac{\partial g}{\partial E}-\frac{1-B \Lambda}{E B}\frac{\partial g}{\partial \Lambda} \right)
\end{aligned} 
\end{equation}
Then the collisional operator is 
\begin{equation}
\begin{aligned}
     \Gamma&_a^{-1} \left(\frac{\partial f_a}{\partial t}\right)_c= -\frac{v_\parallel}{BE}\frac{\partial}{\partial E}\left( \frac{2BE^2}{v_\parallel} f\frac{\partial h}{\partial E} \right) - \frac{v_\parallel}{BE}\frac{\partial}{\partial \Lambda} \left[ \frac{2\Lambda(1-B\Lambda)}{v_\parallel} f \frac{\partial h}{\partial \Lambda}\right] \\
    &+\frac{1}{2} \frac{v_\parallel}{BE} \Bigg\{ \frac{\partial ^2}{\partial E^2} \left[
    \frac{BE}{v_\parallel}f\left(4E^2\frac{\partial^2 g}{\partial E^2}+2E\frac{\partial g}{\partial E} \right) \right] \\
    &+2\times\frac{\partial^2}{\partial E\partial\Lambda}\left[ \frac{BE}{v_\parallel} f\frac{4\Lambda(1-B\Lambda)}{B} \left( \frac{\partial^2 g}{\partial E\partial \Lambda} -\frac{1}{2E}\frac{\partial g}{\partial \Lambda}\right)  \right] \\
    &+\frac{\partial^2}{\partial \Lambda^2}
    \frac{BE}{v_\parallel} f\left[ \frac{2\Lambda(1-B \Lambda )}{BE}\right]^2 \left[\frac{\partial^2 g}{\partial \Lambda^2 }+\frac{BE}{2\Lambda(1-B \Lambda )}\frac{\partial g}{\partial E} - \frac{2\Lambda B -1}{2\Lambda(1-B \Lambda )}\frac{\partial g}{\partial \Lambda} \right] \Bigg\} \\
    &+\frac{1}{2} \frac{v_\parallel}{BE} \Bigg\{ \frac{\partial}{\partial E}\frac{BE}{v_\parallel}f \left[ -2E \frac{\partial ^2 g}{\partial E^2}-\frac{2\Lambda(1-B\Lambda)}{BE}\frac{\partial^2 g}{\partial \Lambda^2}-3\frac{\partial g}{\partial E} +\frac{\Lambda}{E}\frac{\partial g}{\partial \Lambda} \right]\\
    &+\frac{\partial }{\partial \Lambda} \frac{BE}{v_\parallel} f \left[\frac{4\Lambda(1-B\Lambda)}{BE}\frac{\partial^2 g}{\partial E\partial\Lambda}+\frac{2\Lambda(1-B\Lambda)(2B\Lambda-1)}{(BE)^2}\frac{\partial^2 g}{\partial \Lambda^2} -\frac{2+B\Lambda(3B\Lambda-4)}{(BE)^2}\frac{\partial g}{\partial\Lambda}+\frac{\Lambda}{E}\frac{\partial g}{\partial E} \right] \Bigg\}  \\
\end{aligned}
\end{equation}
For isotropic distribution, $g(v)$ and $h(v)$ are independent of $\Lambda$. Then we have,
\begin{equation}
\begin{aligned}
     \Gamma_a^{-1} \left(\frac{\partial f_a}{\partial t}\right)_c & = -\frac{v_\parallel}{BE}\frac{\partial}{\partial E}\left( \frac{2BE^2}{v_\parallel} f\frac{\partial h}{\partial E} \right) \\
    &+\frac{1}{2} \frac{v_\parallel}{BE} \Bigg\{ \frac{\partial ^2}{\partial E^2} \left[
    \frac{BE}{v_\parallel}f\left(4E^2\frac{\partial^2 g}{\partial E^2}+2E\frac{\partial g}{\partial E} \right) \right] +\frac{\partial^2}{\partial \Lambda^2}
    \frac{BE}{v_\parallel} f\left[ \frac{BE}{2\Lambda(1-B \Lambda )}\frac{\partial g}{\partial E} \right] \Bigg\} \\
    &+\frac{1}{2} \frac{v_\parallel}{BE} \Bigg\{ \frac{\partial}{\partial E} \left[ \frac{BE}{v_\parallel}f \left( -2E \frac{\partial ^2 g}{\partial E^2}-3\frac{\partial g}{\partial E} \right)\right] +\frac{\partial }{\partial \Lambda} \frac{BE}{v_\parallel} f \left( \frac{\Lambda}{E}\frac{\partial g}{\partial E} \right) \Bigg\}.  \\
\end{aligned}
\end{equation}
The off-diagonal diffusion terms $\partial^2/\partial E\partial \Lambda$ do not exist. There are singularities when $\Lambda=0$ and $\Lambda=1$. The collision operator derived from Rosenbluth potential will be further studied and reported in a separate work \cite{lu2024high}.
\section*{Acknowledgments}
The authors would like to thank Dr. F. Zonca, Dr. Hogun Jhang, Dr. S.D. Pinches, Dr. A. Chankin, Dr. Baolong Hao, Dr. G. Tardini for fruitful discussions, partially within the EUROFUSION Enabling Research Projects Projects ``ATEP''. This work has been carried out within the framework of the EUROfusion Consortium, funded by the European Union via the Euratom Research and Training Programme (Grant Agreement No 101052200 — EUROfusion). Views and opinions expressed are however those of the author(s) only and do not necessarily reflect those of the European Union or the European Commission. Neither the European Union nor the European Commission can be held responsible for them. ITER is the Nuclear Facility INB no. 174. The views and opinions expressed herein do not necessarily reflect those of the ITER Organisation.

\section*{References}
\bibliographystyle{unsrt}
\bibliography{collision}
%
%
%

\end{document}